\documentclass[a4paper]{jpconf}
\usepackage{graphicx}
\usepackage[latin1]{inputenc}
\begin{document}
\title{Point Contact Spectra on YBa$_2$Cu$_3$O$_{7-x}$/La$_{0.7}$Ca$_{0.3}$MnO$_3$ bilayers}

\author{S. PIANO, F. BOBBA, A. DE SANTIS, F. GIUBILEO, \\ A. SCARFATO and
A. M. CUCOLO}

\address{Physics Department and CNR-SUPERMAT Laboratory,
University of Salerno,\\ Via S. Allende, 84081 Baronissi (SA),
Italy}

\ead{samanta@sa.infn.it}

\begin{abstract}
We present conductance characteristics of point contact junctions
realized between a normal Pt-Ir tip and
YBa$_2$Cu$_3$O$_{7-x}$/La$_{0.7}$Ca$_{0.3}$MnO$_3$ (YBCO/LCMO)
bilayers. The point contact characteristics show a zero bias
conductance peak, as a consequence of the formation of Andreev bound
states at the YBCO Fermi level. The temperature evolution of the
spectra reveals a depressed zero bias peak and a reduced
superconducting energy gap, both explainable in terms of spin
polarization effects due to the LCMO layer.
\end{abstract}

\section{Introduction}

The  investigation of  hybrid Superconducting/Ferromagnetic
heterostructures aims at understanding the mechanism of electronic
transport  in the presence of competing superconducting and
ferromagnetic orders. This research has also potential technological
outlets in the context of spintronics \cite{FSAppl}. A major step in
the field has been obtained with the discovery of perovskite
manganites which exhibit colossal magnetoresistance (CMR)
\cite{CMR}. The possibility of engineering hybrid structures based
on these ferromagnetic oxide compounds and high-Tc superconducting
materials, opens up appealing perspectives. At the very heart lies
the fundamental interest for their physical properties, such as the
modification of the density of states (DOS) of the superconducting
samples  due to the effects of the magnetic layer \cite{PRL}.
 Soulen \emph{et al.} \cite{Soulen} have shown
that the Point Contact Andreev Reflection can be used to determine
the spin polarization of the ferromagnetic materials. In fact in the
presence of a ferromagnetic material the Andreev Reflection
probability at the Superconductor/Ferromagnetic (S/F) interface is
reduced by the carrier density of the minority spin band at the
Fermi level.  Many authors have measured the polarization of
different ferromagnetic materials using this technique
\cite{Strikers}.

In this work we analyze the effect of  spin-polarized electrons on
the tunneling current in a heterostructure constituted by a high-Tc
superconductor (YBCO) and a CMR ferromagnetic oxide (LCMO). We
observe the presence of both Andreev bound states in the YBCO layer,
and spin polarization in the LCMO layer. The zero bias conductance
peak, appearing in the conductance spectra due to Andreev bound
states at the Fermi level of the superconductor, results to be
depressed by a proximity effect induced by the magnetic layer. Our
results are well interpreted in the framework of the spin-polarized
transport theory.

\section{Samples preparation and experimental setup}

\begin{figure}[t!]
\includegraphics[width=12pc,height=11pc]{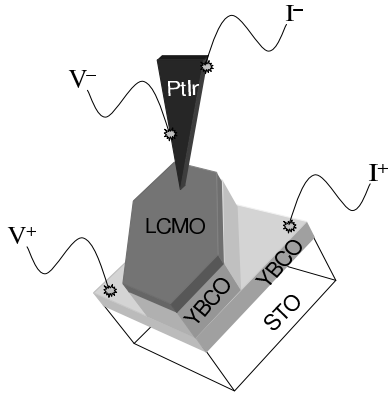}\hspace{2pc}%
\begin{minipage}[b]{24pc}
\caption{\label{scheme} Scheme of our Point Contact junction
realized between a Pt-Ir tip and a YBCO/LCMO bilayer. \\
\\}
\end{minipage}
\end{figure}

Highly epitaxial, \emph{c}-axis oriented
YBa$_2$Cu$_3$O$_{7-x}$/La$_{0.7}$Ca$_{0.3}$MnO$_3$ (YBCO/LCMO)
heterostructures were grown in a pure oxygen atmosphere (p=3.0 mbar)
on SrTiO$_3$ (0 0 1) substrates (STO) by DC sputtering technique at
T=900°C (for further details see Refs. \cite{Adele, Adele_tesi}). A
YBCO film of $500${\AA} was first deposited; then, by defining the
geometry through a shadow mask, we sequentially realized a YBCO/LCMO
bilayer with thicknesses  $d_{\rm YBCO}=1000${\AA} and $d_{\rm
LCMO}=75${\AA}, respectively.
 The conductance spectra were measured by using a home-built Point Contact Andreev
 Reflection (PCAR) setup operating from liquid-helium temperature to room temperature.
 To realize the Point Contact experiments, we used mechanically cut
  fine tips of Pt-Ir, chemically etched in a
$40\%$ solution of HCl in an ultrasound bath. Samples and tips were
placed in the PCAR probe and the electrical contacts (two on the
tip, and two on the first YBCO basis) were realized by indium drops.
In  Fig. \ref{scheme} we show the geometry of the junction and
 the voltage-current terminals. At low temperature, we established the contact between the Pt-Ir tip
and the YBCO/LCMO bilayer  using a micrometric screw.
 The current-voltage  ($I$ vs $V$) characteristics were measured by using a conventional four-probe
 method. A lock-in technique with an ac current of amplitude less than $1 \mu A$
 was used to measure the differential conductance
 ($dI/dV$ vs $V$) spectra as functions of the applied voltage.

\section{Results and discussion}

\begin{figure}[t!]
\includegraphics[width=20pc]{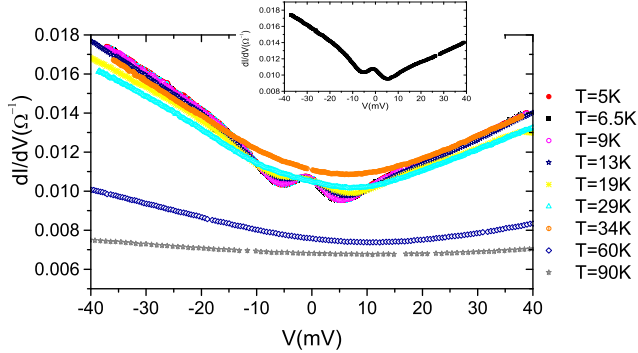}\hspace{2pc}%
\begin{minipage}[b]{16pc}\caption{\label{YBCO_LCMO} Temperature dependence of a highly stable junction
between the YBCO/LCMO bilayer and the Pt-Ir tip. We notice that a
ZBCP appears at low temperatures and disappears at about 30K. The
lowest-temperature spectrum is detailed in the inset. \\
\\}
\end{minipage}
\end{figure}

In this section we present the conductance spectra obtained between
a YBCO/LCMO bilayer and a Pt-Ir tip. We remark that our transport
measurements include two different interfaces, YBCO/LCMO and
LCMO/Pt-Ir, both responsible for the profile of the conductance
curves. The different contributions can be evidenced for instance in
the lowest-temperature spectrum (see the inset of Fig.
\ref{YBCO_LCMO}), characterized by an asymmetric, ``V''-shaped
background, with the presence of a Zero Bias Conductance Peak
(ZBCP). The ``V''-shaped background, similar to that reported for
other metallic oxide junctions \cite{oxide1,oxide2},  is a signature
of the LCMO/Pt-Ir junction, while the YBCO layer is responsible for
the asymmetry \cite{Deutscher}. On the other hand, the ZBCP is a
consequence of the \emph{d-wave} symmetry of the superconducting
order parameter of YBCO, indicating the formation of the Andreev
bound states at the YBCO Fermi level
\cite{Deutscher,YBCO,Tanaka_dwave}. Moreover, the presence of a ZBCP
suggests that our tunnel junction is not completely \emph{c}-axis
oriented, but a component in the \emph{a-b} plane is present as
well.

The nature of the ZBCP can be experimentally investigated by
following the temperature evolution of the conductance spectra. In
the case of PCAR on pure YBCO, the literature reports a ZBCP
decreasing with increasing temperature, and vanishing at the
critical temperature of YBCO ($\sim$ 90 K)
 \cite{Deutscher}. In our YBCO/LCMO junction (see Fig. \ref{YBCO_LCMO}),
 we observe instead a depressed ZBCP. According to the theory of spin transport between
a ferromagnetic material and a  $d$-wave superconductor, the
depression of the ZBCP follows from the suppression of Andreev
reflections at the interface, due to the spin polarization of the
ferromagnetic layer \cite{supposta}. In the measured conductance
spectra, the ZBCP disappears at a temperature of about $T_c \sim$ 30
K, which is in agreement with the resistivity measurements on
YBCO/LCMO bilayers \cite{Adele_tesi}. This fact provides further
evidence that the ZBCP is a consequence of the superconducting
nature of YBCO and is not due to spurious effects like inelastic
tunneling via localized magnetic moments in the barrier region
\cite{Cucolo}.

From our conductance spectra, the amplitude of the superconducting
order parameter of YBCO can be inferred as well. Namely, we can fit
the background according to the model \cite{libro,YBCOLCMO}:
\begin{equation}\label{LCMOback}
G(V)=\frac{dI}{dV}\propto \frac{d}{dV}\int
N_{FM}(E)N_{SC}(E+eV)[f(E)-f(E+eV)]dE,
\end{equation}
where $N_{SC}$ is the DOS of the YBCO layer, and $N_{FM}$ is the DOS
of the LCMO layer. The latter can be expressed as
$N_{FM}(E)=N_{FM}(0)[1+(|E|/\Lambda)^{n}]$, where  $\Lambda$ is a
constant associated with the electron correlated energy of LCMO at
the interface and the exponent $0.5 < n < 1$ reflects the degree of
disorder in LCMO near the YBCO/LCMO interface. Concerning the YBCO,
we can assume that, for bias voltages larger than the
superconducting energy gap, $N_{SC}$ is approximately constant,
except for a linear correction taking into account for the asymmetry
of the normal state of YBCO. In a window of bias voltages $V \in
[-\bar V, \ \bar V]$, the DOS of YBCO can thus be written as $N_{SC}
= 1 - \kappa (V + \bar V)$, where $\kappa$ is the asymmetry factor
and the total conductance is normalized such that $G(-\bar V)=1$.

\begin{figure}[b!]
\includegraphics[width=18pc]{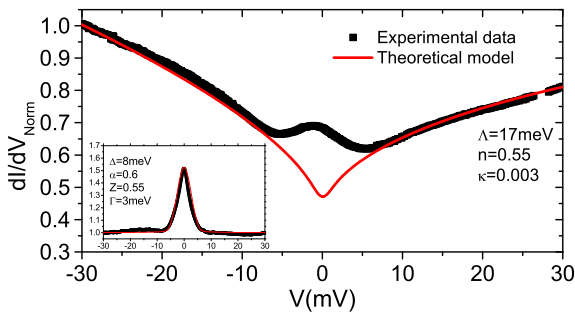}\hspace{2pc}%
\begin{minipage}[b]{18pc}\caption{\label{background} The conductance spectrum with the fitted background. In the inset
the YBCO conductance curve is shown with the best fit with
\emph{d-wave} symmetry of the order parameter. \newline \newline
\newline}
\end{minipage}
\end{figure}

In Fig. \ref{background}, we show the lowest-temperature conductance
spectrum, together with the best fitting curve for the background,
following Eq. (\ref{LCMOback}). The superconducting energy gap of
YBCO corresponds to the bias voltage at which the theoretical curve
for the background deviates from the experimental conductance
spectrum. From Fig. \ref{background}, we can estimate an energy gap
$\Delta \sim$~8~meV. The contribution of the YBCO layer to the
conductance characteristic can then be obtained dividing the
measured differential conductance (at the lowest temperature) by the
modeled background curve: $G_{\rm YBCO}(V) =
G(V)_{exp}/G(V)_{back}$. We can satisfactorily fit the spectrum
$G_{\rm YBCO}(V)$ with a \emph{d}-wave BTK model
\cite{Tanaka_dwave}, regarding as fitting parameters the
superconducting gap $\Delta$, the barrier strength $Z$, the angle
$\alpha$ of the order parameter and the smearing factor $\Gamma$.
Remarkably (see the inset of Fig. \ref{background}), the fitting
provides a gap $\Delta =$ 8 meV, consistent with our previous
findings. This value, smaller than the reported gap value of 20meV
for YBCO \cite{Deutscher}, can be explained by the proximity effect
of the Cooper pairs from YBCO to LCMO, or by the injection of
spin-polarized electrons from LCMO to YBCO \cite{YBCOLCMO}.

\begin{figure}[t!]
\includegraphics[width=18pc]{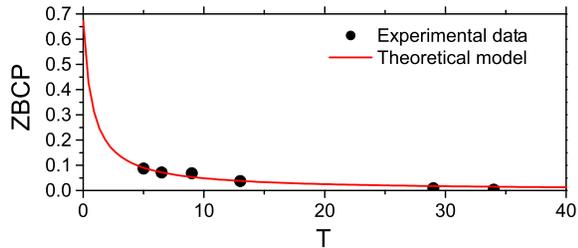}\hspace{2pc}%
\begin{minipage}[b]{18pc}\caption{\label{picco} Temperature dependence of ZBCP (dots)
with the best theoretical fit (solid line). \\ \\ \\
 }
\end{minipage}
\end{figure}

It has been theoretically predicted that, in a S/F interface, the
amplitude of the spin polarization of  the ferromagnetic layer can
be estimated from the temperature dependence of the ZBCP. The latter
is expected to be proportional to the inverse of the temperature for
intermediate temperatures \cite{Tanaka}. In Fig. \ref{picco} we show
the temperature dependence of the ZBCP as obtained by the
experimental conductance spectra of Fig. \ref{YBCO_LCMO}, together
with the best fitting curve. With an extrapolation of the best
fitting curve at low temperatures ($T \rightarrow 0$), we can
directly compare our ZBCP evolution to the theoretical model of Ref.
\cite{Tanaka}. This allows us to estimate a spin polarization of the
LCMO layer of about $67\%$, meaning that the electron spins are not
fully polarized. This is consistent with the observation of a
(depressed) ZBCP: in fact, it would have been completely suppressed
if the LCMO polarization had approached unity.

In conclusion, PCAR measurements  on the heterostructure constituted
by a YBCO/LCMO bilayer and a Pt-Ir tip have resulted in conductance
spectra with
 a ZBCP, signature of the \emph{d}-wave symmetry of
the order parameter of YBCO. A reduced height of the ZBCP, as
compared to the case of superconductor/normal metal junctions, has
been observed. We have shown that this effect is due to a partial
polarization of the LCMO layer.

A deeper analysis of the effects of one or more magnetic layers on
the superconducting properties of high-Tc compounds awaits further
investigation.

\section*{References}

\smallskip


\begin{thebibliography}{99}

\bibitem{FSAppl} Prinz G A 1995 {\it Phys. Today} {\bf 48} 58; Prinz G A 1998 {\it Science} {\bf 282}
1660.

\bibitem{CMR}
{\em Colossal Magneto-resistive Oxides} 2000 edited by
Tokura Y (Gordon and Breach Science, Netherlands).


\bibitem{PRL} Pe${\rm{\tilde{n}}}$a V \emph{et al.} 2005 {\it Phys. Rev. Lett.}
{\bf 94} 057002.
\bibitem{Soulen} Soulen R J  \emph{et al.} 1998 {\it Science} {\bf 282}
85.
\bibitem{Strikers} Strijkers G J  \emph{et al.} 2001 {\it Phys. Rev. B} {\bf
63};  Chien Y L,  Tomioka Y and Tokura Y  2002 {\it ibid.} {\bf 66}
012410;  Panguluri R P  \emph{et al.} 2003 {\it ibid.} {\bf 68}
201307.


\bibitem{Adele} De Santis A  \emph{et al.}  2004 {\it Physica C} {\bf
408} 48.
\bibitem{Adele_tesi} De Santis A  2003 {\it PhD thesis}  (University of Salerno).

\bibitem{oxide1} Raychaudhuri A K  \emph{et al.} 1995 {\it Phys. Rev.
B} {\bf 51} 7421.
\bibitem{oxide2} Cao D  \emph{et al.} 2000 {\it Phys. Rev. B} {\bf
61} 11 373.

\bibitem{Deutscher} Deutscher G 2005 {\it Rev.
Mod. Phys.} {\bf 77} 109.

\bibitem{YBCO} Hu C-R 1994 {\it Phys. Rev. Lett.} {\bf 72} 1526; Wei J Y T \emph{et al.} 1998 {\it ibid.}   {\bf 81}
2542.

\bibitem {Tanaka_dwave} Kashiwaya S and Tanaka Y 2000 {\it Rep. Prog. Phys.} {\bf 63}
1641.

\bibitem{supposta} Zhu J-X, Friedman B and Ting C S 1999  {\it Phys. Rev. B} {\bf 59} 9558;
Kashiwaya S \emph{et al.}  1999 {\it ibid.} {\bf 60} 3572.

\bibitem{Cucolo} Cucolo A M 1998 {\em  Physica C} {\bf 305} 85.

\bibitem{libro} Altshuler B and  Aronov A 1985 in {\it Electron-Electron Interactions in Disordered
Systems}
 edited by Efros A L and Pollak M (North-Holland, Amsterdam).

\bibitem{YBCOLCMO} Luo P S  \emph{et al.} 2005 {\it Phys. Rev. B} {\bf 71} 094502.

\bibitem{Sawa} Sawa A  \emph{et al.} 2000 {\it Physica C} {\bf 339}
287.


\bibitem{Tanaka} Hirai T, \emph{et al.} 2003 {\it Phys. Rev. B} {\bf 67}
174501.



\end{thebibliography}
\end{document}